\documentclass[prb,aps,reprint,amsmath,amssymb,twocolumn]{revtex4-1}
\usepackage{amsmath}
\usepackage{graphicx}
\usepackage{epsfig} 
\usepackage{textcomp}
\usepackage{amsfonts}
\usepackage{color}
\usepackage{natbib}

\begin{document}

\title{Breakdown of Stokes-Einstein relation in the supercooled liquid state of phase change materials}
\author{Gabriele C. Sosso$^{1}$}
\author{J\"{o}rg Behler$^{2}$}
\author{Marco Bernasconi$^{1}$}\email[Corresponding author. Email address: ]{marco.bernasconi@unimib.it}

\affiliation{
$^1$ Dipartimento di Scienza dei Materiali, Universit\`{a} di Milano-Bicocca,
Via R. Cozzi 53, I-20125 Milano, Italy}

\affiliation{
$^2$ Lehrstuhl f\"{u}r Theoretische Chemie, Ruhr-Universit\"{a}t Bochum, Universit\"{a}tsstrasse 150, D-44780 Bochum, Germany}


\begin{abstract}
The application of amorphous chalcogenide alloys as data-storage media relies on their ability to undergo an extremely fast (10-100 ns)
crystallization once heated at sufficiently high temperature.
However, the peculiar features that make these materials so attractive for memory devices still lack a comprehensive microscopic understanding.
By means of large scale molecular dynamics simulations, we demonstrate that the supercooled liquid of the prototypical compound GeTe shows a very high atomic mobility ($D$ $\sim$ 10$^{-6}$ cm$^2$/s) down to temperatures close to the glass transition temperatures. This behavior
leads to a breakdown of the Stokes-Einstein relation between the self-diffusion coefficient and the viscosity in the supercooled liquid. The results suggest that the fragility of the supercooled liquid is the key to understand the fast crystallization process in this class of materials.
\end{abstract}

\maketitle   

\section{Introduction}
Phase-change materials based on chalcogenide alloys are attracting a lot of interest
 due to their ability to undergo reversible and fast transitions between the amorphous
and crystalline phases upon heating \cite{wuttig07,pirovano,lacaita,lencerReview,welrev}. This property is exploited in rewriteable optical
media (DVD, Blu-Ray Discs) and phase change non volatile memories (PCM).
The strong optical and electronic contrast between the crystal and the amorphous allows discriminating
between the two phases that correspond
to the two states of the memory.
The PCM devices, first proposed by Ovshinsky  in the late 1960's \cite{ovshinsky},
 offer extremely fast programming, extended cycling endurance, good reliability and inexpensive,
 easy integration.
A PCM is essentially a resistor of a thin film of a chalcogenide alloy
(typically Ge$_2$Sb$_2$Te$_5$, GST) with a low field resistance that changes by several orders of magnitude across the phase change.
 In  memory operations,
 cell read out is performed at low bias. Programming the memory requires instead a relatively large current to heat
up the chalcogenide and induce the phase change, either the melting of the crystal and subsequent amorphization (reset) or the
 recrystallization of the amorphous (set).

The key property that makes these materials suitable for applications in  PCM
is the high speed of the transformation which leads to  full
crystallization on the time scale of 10-100 ns upon Joule heating.
What makes some chalcogenide alloys so special in this respect and so different from most amorphous semiconductors
is, however, still a matter of debate.

In this paper, we demonstrate by means of atomistic
molecular dynamics (MD) simulations that the phase change compound GeTe  displays  a very high mobility in the supercooled liquid phase
down to very low temperatures $T$
leading to a breakdown of  the Stokes-Einstein relation
$D \propto \frac{T}{\eta}$ (SER) between diffusivity $D$ and viscosity $\eta$.
The breakdown of SER in the supercooled liquid phase, which is typical of a fragile liquid \cite{VTF},
 is actually the key to understand the
fast crystallization behavior of these materials as discussed below.

Supercooled liquids are classified as fragile or strong  on the basis of the temperature dependence of their viscosity
\cite{VTF}.
An ideal strong liquid shows
an Arrhenius  behavior  of the viscosity $\eta$ from the melting temperature T$_m$ down to the glass transition temperature T$_g$.
On the contrary,
in a fragile liquid  $\eta$ follows an Arrhenius behavior
only  above a cross-over temperature T$^*$  below which a Vogel-Tammann-Fulcher (VTF) function $\eta=\eta_o \exp(\frac{E}{k_B (T-T_o)})$
is customarily used to reproduce the data with  $\eta_o$, $E$ and T$_o$ as fitting parameters \cite{VTF}.
The viscosity in a fragile liquid shows a steep rise by approaching T$_g$ which according to the SER
would lead to a strong decrease in the atomic mobility.
On the other hand, the self diffusion coefficient $D$ controls both the speed of crystal growth $u$ and the steady state nucleation rate
$I_{ss}$. In fact,
 classical nucleation theory \cite{nucl,welrev} predicts that $u \propto D (1-\exp(-\Delta G/(k_B T))$
 and  $I_{ss} \propto D   \exp(-G_c/(k_B T))$ where
$\Delta G$ is the free energy difference between
the liquid (or amorphous) and the crystalline phases
and $G_c$ is the formation free energy of the critical nucleus  given in turn
by $G_c= 16 \pi\sigma^3/(3\Delta G^2)$ with $\sigma$  interface energy.
$\Delta G$ is the driving force for crystallization that
decreases by increasing temperature and finally vanishes at  T$_m$.
In phase change materials, due to the breakdown of SER, the diffusivity can be
very high just above T$_g$ in spite of a large viscosity. Consequently
$D$ can reach high values  at temperatures  much lower than T$_m$ where a large
driving force for crystallization is present.

A breakdown of SER in GST has actually been  suggested by
recent   measurements of the crystallization
rate by Orava et al. \cite{orava}.
Still, further evidences of the breakdown of SER are needed
to support the crucial assumptions made by
Orava {\sl et al.} to infer the viscosity from their only data
of differential scanning calorimetry.

To this aim we performed MD simulations by using
a classical interatomic potential  \cite{NNGeTe} we generated by fitting
a large database of density functional energies by means of the
Neural Network (NN) method introduced in Ref. \cite{behler}.
So far we have restricted ourselves to the binary compound GeTe which is also under scrutiny for applications in PCM and
which shares most of the properties with the more commonly used GST \cite{lencerReview,akola,mazza}.
The NN potential displays an accuracy close to that of the underlying density functional theory (DFT) framework
 at a much reduced computational load that scales linearly with the size of the system.
It allows us  to simulate several thousands of  atoms for tens of ns, which is well beyond present-day capabilities of DFT molecular dynamics.

\section{Computational details}
The NN interatomic potential of GeTe was obtained in Ref. \cite{NNGeTe} by fitting a huge database of DFT energies by means of the method introduced by Behler and Parrinello \cite{behler}.   The database consists of the total energies of about 30000 configurations of 64-, 96-, and 216-atom supercells computed by employing the the Perdew-Burke-Ernzerhof (PBE) exchange and correlation functional \cite{PBE} and norm conserving pseudopotentials. The NN potential displays an accuracy close to that of the underlying DFT-PBE framework whose reliability in describing structural and dynamical properties of GeTe and other phase change materials has been validated in several previous works  by our \cite{NNGeTe,mazza,apl} and other groups \cite{akola}.

The simulations were performed with the NN code RuNNer~\cite{RuNNer} and a 4096-atom cubic supercell
by using  the DL$\_$POLY~\cite{dlpoly} code as MD driver.
The time step  was  set to 2~fs, and constant temperature was enforced
by a stochastic thermostat~\cite{bussi}.

It turned out that in order to reproduce the equilibrium density of the liquid at T$_m$,
an empirical van der Waals (vdW) correction had to be added to the NN
potential. This was done by using the scheme proposed by Grimme \cite{grimme} with  the $s_6$ parameter tuned to
0.55 to reproduce the experimental equilibrium volume of the liquid at T$_m$\cite{expliquid}.
The experimental equilibrium volume of the amorphous and crystalline phases are instead well reproduced by the NN potential
without the need of the vdW interaction \cite{NNGeTe}.
The inability of the NN potential in reproducing the equilibrium volume of the liquid can be traced back to the
presence of nanovoids in the melt \cite{akola}.
In the liquid  the nanovoids can coalesce and increase in size by decreasing the density
which results into a reduced tensile stress upon expansion.
This effect is hindered  by  vdW interactions.
Nanovoids are also present in the amorphous phase  \cite{akola},
 but their distribution can not change with temperature because of the low
atomic mobility in the amorphous phase.
The calculated linear thermal expansion coefficient ($\alpha$=$\frac{1}{3 V}\frac{\partial V}{\partial T}$) of the liquid at T$_m$ with the NN+vdW potential
 turned out to be 4.73 $\cdot$ 10$^{-5}$ K$^{-1}$  to be compared with the experimental value of
 3.73 $\cdot$ 10$^{-5}$ K$^{-1}$ \cite{expliquid}.

The added vdW interaction acts just as a
volume dependent term in the equation of state of the liquid but it is not included in the MD simulations discussed below.
DFT calculations of the self-diffusion coefficient at two selected temperatures
were performed by molecular dynamics simulations within the Born-Oppenheimer approach by using  the code CP2K~\cite{cp2k};
  Kohn-Sham orbitals were expanded in a Triple-Zeta-Valence plus Polarization
 Gaussian-type basis set  and the charge density was expanded
in a planewave basis set with a cut-off of 100~Ry to efficiently
solve the Poisson equation within the Quickstep
scheme \cite{cp2k}.
 Goedecker-type
pseudopotentials \cite{GTH} with  four and six valence electrons
were used for  Ge and Te, respectively.
Brillouin Zone  integration was
restricted to the  $\Gamma$ point of the 216-atom supercell.
The same framework was used in our previous DFT molecular dynamics
simulations of GeTe \cite{NNGeTe,mazza}.

\section{Results}

To study the properties of the supercooled liquid, we first assessed the ability of the NN potential, and thus of the
underlying DFT-PBE framework, to reproduce T$_m$.
The melting temperature was computed
by means of thermodynamic integration \cite{thermointeg}  that yielded T$_m$=1001 K very close to the experimental
value at normal pressure of 998 K \cite{Tmelt}.
To obtain T$_m$, we first computed the difference in the Helmholtz free energy $F$ between the NN system
and a reference system for which an analytic expression for $F$ is known, at a given temperature $T'$ and density $\rho'$. Namely
\begin{equation}
F_{NN}(T',\rho') -F_{ref}(T',\rho')= \int_0^1 d\lambda \langle U(\lambda) \rangle,
\end{equation}
 where
the average is taken over a MD simulation with the potential $U(\lambda)=\lambda U_{NN} - (\lambda-1)U_{ref}$.
The temperature and density were set to   the experimental values
at the melting point at normal conditions \cite{Tmelt}.
The reference system was chosen as an Einstein crystal for the solid and a Lennard-Jones fluid \cite{LJ} for the liquid.
In the next step, the chemical potentials were evaluated by
integrating the free energy as a function of density starting
from $\rho'$ \cite{ghiringhelli}

\begin{equation}
\begin{array}{c}
\mu_{NN}(T',\rho)=\frac{1}{N}F_{NN}(T',\rho')  +
b(T')(ln\frac{\rho}{\rho'}+1) + \\ +  \frac{a(T')}{\rho'} + c(T')(2\rho-\rho')
\end{array}
\end{equation}

where $N$ is the number of particles in the simulation cell, and the parameters $a(T')$, $b(T')$ and $c(T')$ were obtained by fitting
the pressure dependence of the density using
\begin{equation}
P(T',\rho)=a(T')+b(T')\rho+c(T')\rho^2.
\end{equation}
 By equating the chemical potential of the two phases one obtains a
transition pressure of -0.44 GPa at the chosen temperature T$'$=998 K. From the calculated Clausius-Clapeyron equation ($dT/dP$=6.85 K/GPa
from the calculated $\Delta S$=$\Delta E/T$ and $\Delta V$ on the theoretical melting line at T=998 K) we then obtained the theoretical melting temperature at normal pressure which is T$_m$=1001 K.

We then analyzed the properties of the supercooled liquid below T$_m$ by computing independently   $\eta$ and
$D$ in microcanonical MD  simulations. The volume of the supercooled liquid was scaled with temperature according to the thermal
expansion coefficient ($\alpha$ is little dependent on temperature) given in Sec. II.
We scaled the temperature from 1000 K to 500 K in eleven steps. At each step the system is equilibrated for 25 ps with the thermostat. Overall
the system is thus quenched from 1000 K to 500 K in 250 ps. At each temperature statistical averages are then collected on microcanonical
simulations  up
to 2 ns long for the calculation of the viscosity as discussed below.

\begin{figure*}[htbp!]
\centerline{\includegraphics[width=1.0\columnwidth]{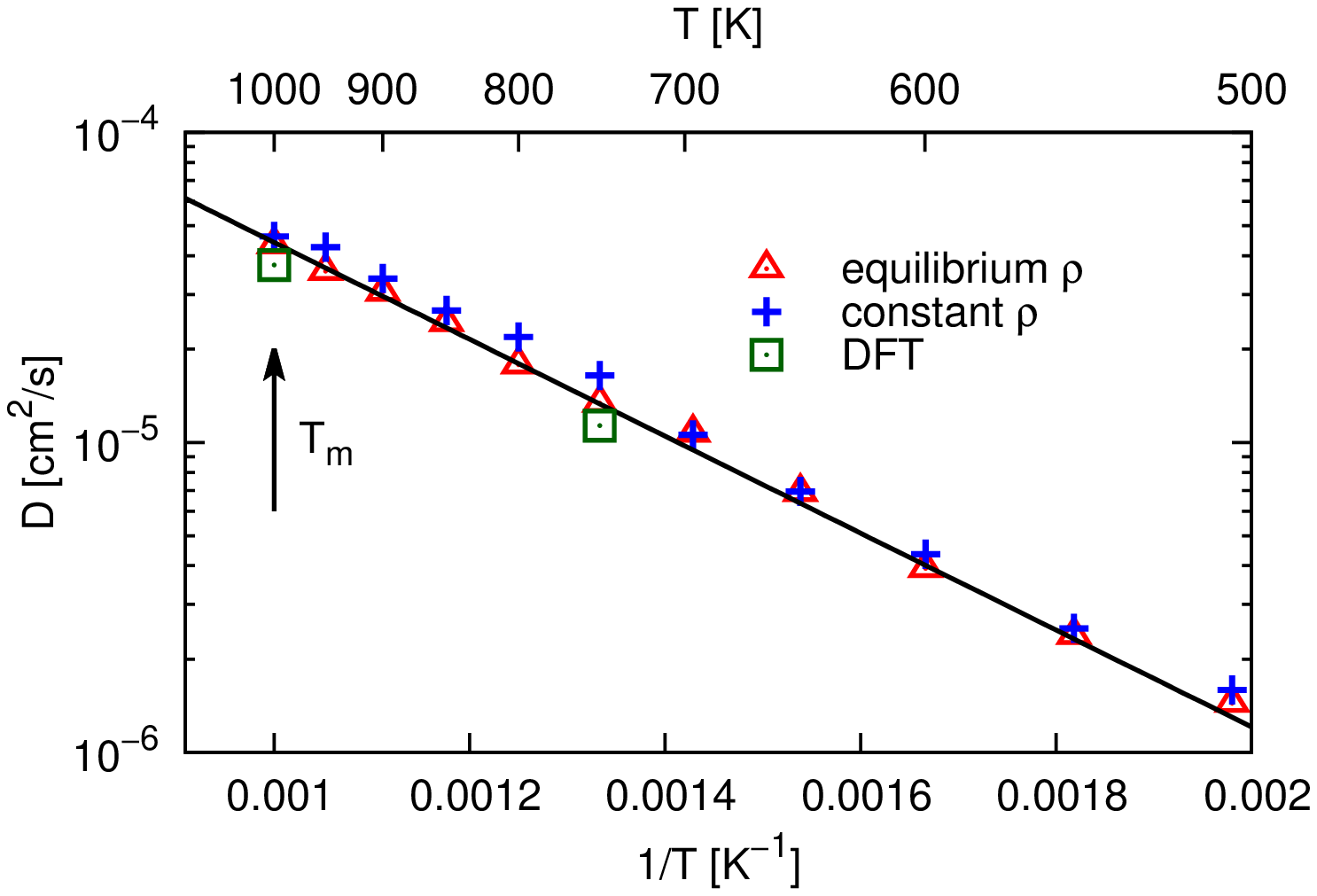}\includegraphics[width=1.0\columnwidth]{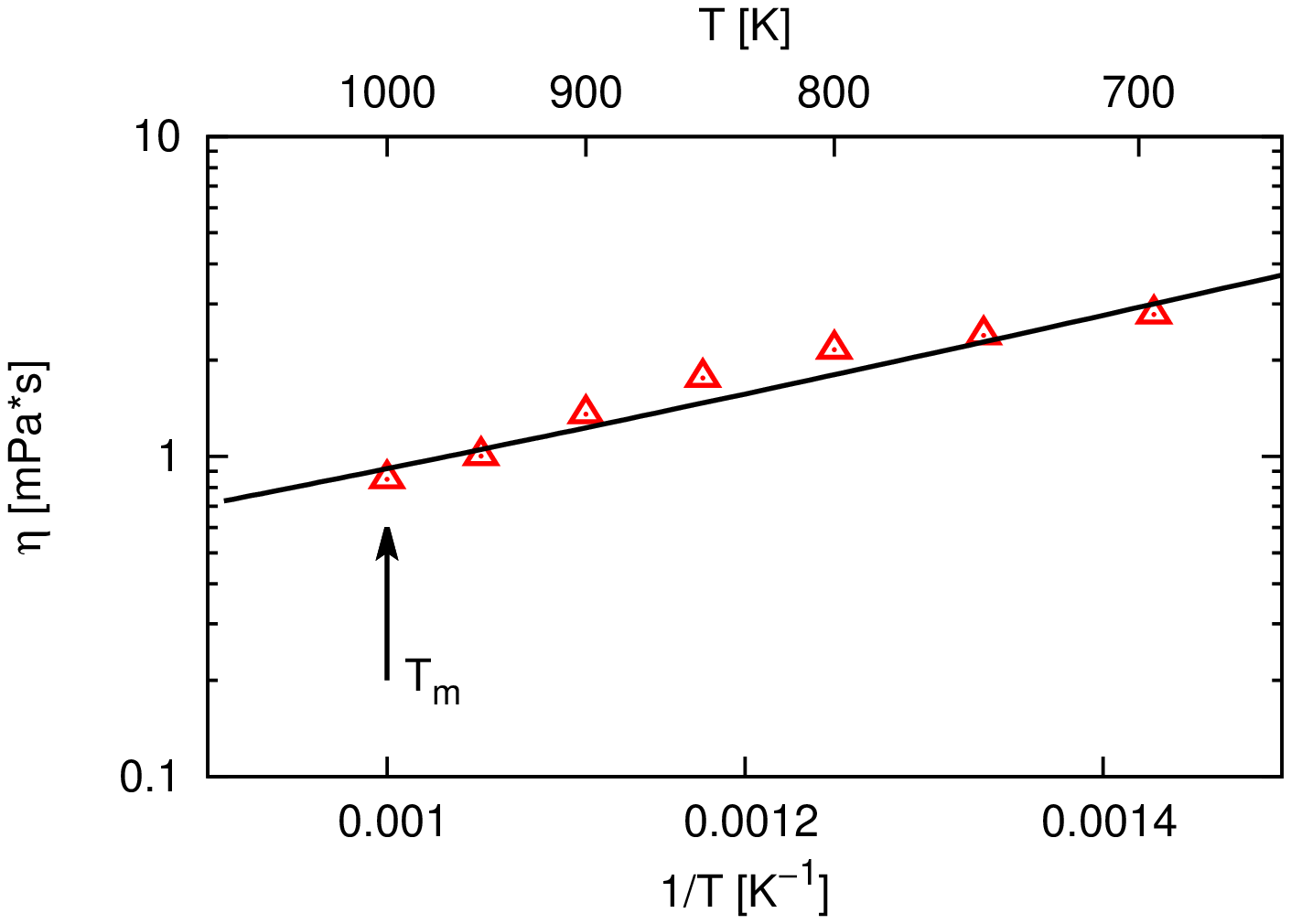}}
\caption{a)  Self-diffusion coefficient $D$ as a function of temperature in the supercooled liquid GeTe
calculated from the mean square displacement (red triangles). The density is scaled with temperature according to the
calculated thermal expansion coefficient.
The open squares are the results of DFT simulations of a 216-atom cell.
The crosses correspond to
the values of $D$ computed by holding the density fixed to the value at the melting point.
b) Calculated viscosity $\eta$ (Green-Kubo formula) as a function of temperature in the supercooled liquid.
The density changes with temperature.
The  lines are Arrhenius fits of the data that give an activation energy of
0.220 $\pm$ 0.002 eV for $D$ and 0.17 $\pm$ 0.035 eV for $\eta$.
T$_m$ is the theoretical melting temperature (see text).
}
\label{arrenhius}
\end{figure*}

\begin{figure}[htbp!]
\centerline{\includegraphics[width=1.0\columnwidth]{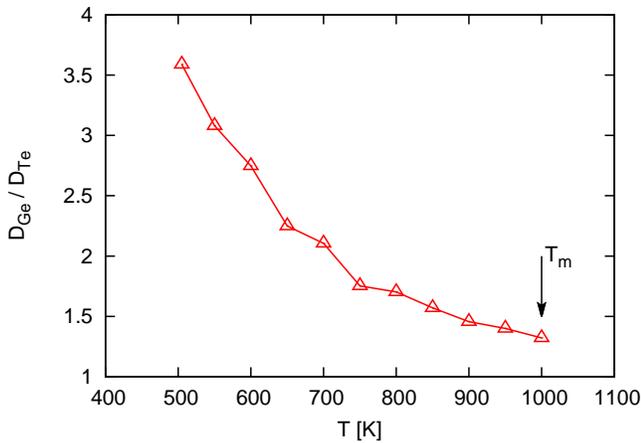}}
\caption{Ratio of the self-diffusion coefficients of the two species $D_{Ge}/D_{Te}$ as a function of temperature
in supercooled liquid GeTe.}
\label{fs0bis}
\end{figure}

We first computed  $D$  from the atomic mean square displacement on the time
scale of 50 ps on which the system does
not crystallize at any temperature we considered. The values of $D$ as a function of temperature are reported in Fig. \ref{arrenhius}a.
$D$ is still of the order of  10$^{-6}$ cm$^2$/s at the lowest temperature of 505 K considered
here; it follows an Arrhenius behavior from T$_m$ to  505 K.
The activation energy  is 0.220 $\pm$ 0.002 eV, a value   much lower than the activation energy  of 1.76 eV obtained from  the Arrhenius dependence
of viscosity in GST measured
in the temperature range  333-373 K (probably  below T$_g$) \cite{otheramorphous}.
The ratio between the self-diffusion coefficient of Ge and Te ($D_{Ge}/D_{Te}$)
increases by decreasing temperature as shown in Fig. \ref{fs0bis}.
The values of $D$ obtained from the NN simulations were also validated by direct DFT molecular dynamics simulations
 at few selected temperatures with a small 216-atom cell
(cf. Fig. \ref{arrenhius}a). The DFT result at 1000 K is equal to that  previously obtained in Ref. \cite{akola} with the same cell and functional
used here.
We checked that the change of volume with temperature has a little effect on the diffusion coefficient as shown in Fig. \ref{arrenhius}a which
also reports the values of $D$ as a function of temperature once the  density is fixed to the value at the melting point.




We then computed $\eta$  between T$_m$ and a temperature T$^*$=700 K which turned out to be
our crossover temperature by means of
 the Green-Kubo (GK) formula \cite{thermointeg}

\begin{equation}
\eta =\frac{V}{k_BT}\int^{\infty}_0 \langle \sigma_{xy}(t)\sigma_{xy}(0) \rangle dt,
\label{GK}
\end{equation}

where $\sigma_{xy}$ is the off diagonal component of the stress tensor and $V$ the supercell volume.
The integral in  Eq. \ref{GK} is converged by restricting the integration time to 60 ps due to the decay of
the self-correlation function above T$^*$. However, long
simulation times up to 2 ns are needed to converge the average ($<..>$)  over different initial times $t=0$.

Above T$^*$ the viscosity can be described by a simple Arrhenius (Fig. \ref{arrenhius}b) function
with an activation energy of 0.17 $\pm$ 0.035 eV, very close to the value of 0.2 eV  measured experimentally
for the Ge$_{0.15}$Te$_{0.85}$ eutectic alloy above T$_m$\cite{eutectic}.
For the GeTe composition, experimental values of $\eta$ are available only at 1000 K
yielding $\eta$=2.59 mPa$\cdot$s which is twice as large as our result (cf. Fig. \ref{arrenhius}b).
This discrepancy is not due to the NN potential but possibly to limitations of the underlying DFT framework.
Previous works on GeSe$_2$ have indeed shown that different choices of the exchange and correlation functional affect the
dynamical properties of the liquid phase \cite{massobrio}.
The viscosity can be computed from the GK formula only above T$^*$ since at lower temperatures the system crystallizes spontaneously
on the time scale of few hundreds of ps which is not long enough to get the value of $\eta$  converged  from Eq. \ref{GK}.
In the supercooled liquid, $\eta$ can not  be defined
on a time scale longer than the crystallization time  which   in GeTe   is  very short in the temperature range 500-700 K.

We thus attempted to extrapolate $\eta$ below T$^*$
by a VTF-like function with the constraint of  matching the  typical
value of 10$^{15}$ mPa$\cdot$ s expected at T$_g$ \cite{VTF}.
Unfortunately, a reliable  value of T$_g$ for is not available from experiments because of the fast crystallization of GeTe,
and its theoretical estimate from simulations is difficult.
Experimental data on T$_g$ are available for the better glass formers Ge$_x$Te$_{(1-x)}$ alloys with $x$=0.15-0.23 \cite{derew}.
By a linear extrapolation with x of these latter data on T$_g$ one obtains T$_g$=511 K for $x$=0.5 which is probably too high.
On the other hand, T$_g$ is customarily assumed to be slightly below the crystallization temperature, that is about 450 K in GeTe \cite{raoux}.
We then used the function proposed in \cite{VTFLarge} that allows fitting $\eta$ over a wider range of temperatures

\begin{equation}
\begin{array}{c}
\log_{10}\eta(T)=\log_{10}\eta_{o}+(15-\log_{10}\eta_{o})\cdot \\
\cdot \frac{T_g}{T} \exp \left [ \left ( \frac{m}{15-\log_{10}\eta_{o}} -1 \right ) \left ( \frac{T_g}{T} -1\right ) \right ]
\label{VTFeq}
\end{array}
\end{equation}

where $m$ and $\eta_o$ are fitting parameters. In Eq. \ref{VTFeq} $\eta$=10$^{15}$ mPa$\cdot$ s at T$_g$.

\begin{figure}[htbp!]
\centerline{\includegraphics[width=1.0\columnwidth]{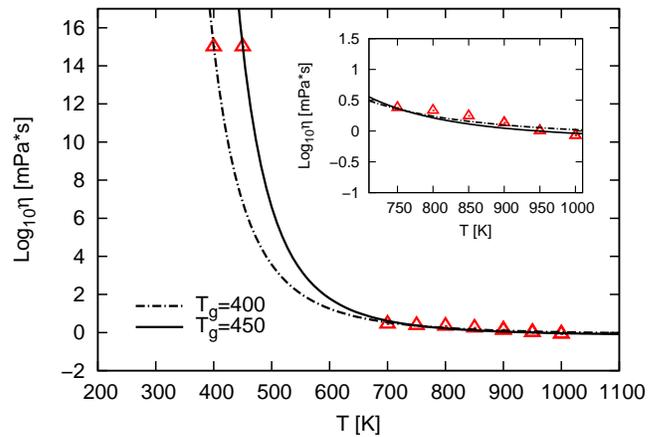}}
\caption{Viscosity  as a function of temperature in supercooled liquid GeTe.
The lines are  fittings with Eq. \ref{VTFeq} of the  Green-Kubo data (triangles, cf. Fig. \ref{arrenhius}b)
above T$^*$ under the constraint of reproducing the value of $\eta$=10$^{12}$ Pa$\cdot$s at T$_g$ with T$_g$=450 K (continuous line) or
T$_g$=400 K (dashed line, see text).
A magnification of the data for T$>$T$^*$ is shown in the inset.}
\label{eta}
\end{figure}

The parameter $m$ is is the fragility index of the supercooled liquid defined by the logarithmic derivative of $\eta$ at T$_g$,
$m=d(log_{10} \eta)/ d(T_g/T)\mid_{T=T_g}$. We performed two fittings for two values of T$_g$ as shown in Fig. \ref{eta}:
the first with T$_g$=450 K
that yields  $m$=111 ($log_{10}\eta_o$=-0.18), and a second one with  a somehow lower temperature
T$_g$=400 K which yields  a very similar value of $m$=104  ($log_{10}\eta_o$=-0.15). Similar results are obtained by using the
modified VFT function proposed in Ref. \cite{VTFnew2}.
Our data on viscosity above T$^*$ are thus consistent with a high fragility of the supercooled liquid.
For sake of comparison we remark that $m$=20 in silica which is a typical strong liquid while $m$=191 in PVC which is a typical fragile liquid \cite{bohmer}.
Unfortunately, due to the lack of data on $\eta$ below T$^*$
and the uncertainties in the value of T$_g$, we can not  assign accurately the degree
of fragility. Nevertheless, even for a value of $m$ as large as 100, the viscosity rises too steeply in the
range 500-600 K to be consistent with the calculated values of $D$ and the application of SER.
In fact, the SER actually breaks down at T $<$ T$^*$ as discussed below.

In the hydrodynamic regime when the SER holds it is actually possible
 to estimate the  viscosity on the shorter time scale  of 50 ps
by a  finite size scaling analysis of the self-diffusion coefficient.
The viscosity can be obtained from
the scaling of $D$ with the edge $L$ of the cubic simulation cell as \cite{artscaling}

\begin{equation}
D(L)= D_{\infty}-\frac{2.387 k_B T}{6 \pi \eta L}.
\label{scaling}
\end{equation}

\begin{figure}[htbp!]
\centerline{\includegraphics[width=1.0\columnwidth]{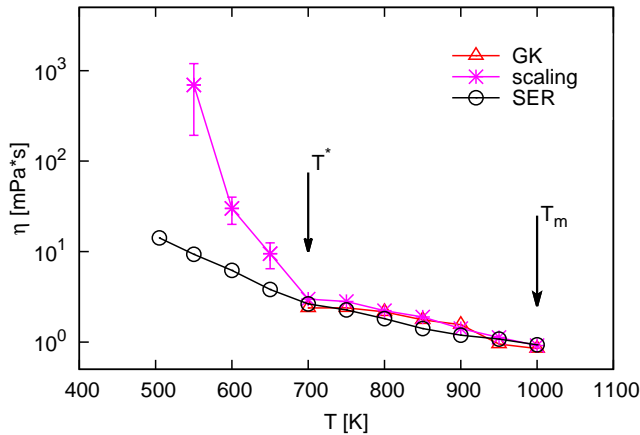}}
\caption{Viscosity computed from the Green-Kubo (GK) formula (red triangles), from the scaling of the diffusion coefficient with the simulation
cell size (stars, Eq. \ref{scaling}), and from the self diffusion coefficient of the 4096-atom cell
and the application of the SER (open circles).}
\label{stokes}
\end{figure}

We considered three cubic models with  512, 1728 or 4096 atoms.
By applying Eq. \ref{scaling} above T$^*$, we obtained values for $\eta$ very close to the
GK data (Fig. \ref{stokes}) and consistent with the SER. However, when Eq. \ref{scaling} is applied below T$^*$,
one obtains values of $\eta$ that are three orders of magnitude larger than those obtained from $D$  and the
application of the SER  ($\eta= \frac{k_B T}{6 \pi R D}$, where $R$ is the average van der Waals radius of the two species) as shown in Fig. \ref{stokes}. This inconsistency  demonstrates that the SER indeed  breaks down.
We remark that the numerical values of $\eta$ reported in Fig. \ref{stokes} below T$^*$ are not
reliable since they are obtained  from Eq. \ref{scaling} which is not applicable when the SER breaks down.

\section{Conclusions}

 We have demonstrated by means of MD simulations that the supercooled liquid of the prototypical phase change compound GeTe shows a  high atomic mobility ($D$ $\sim$ 10$^{-6}$ cm$^2$/s) down to temperatures very close to the glass transition temperature.  Our calculated values of the viscosity as a function of temperature are consistent with a high fragility of the  supercooled liquid.
 However, a compelling assessment of the degree of fragility would require a reliable estimate of T$_g$ which is unfortunately unknown.
 The comparison between the calculated self diffusion coefficient and the viscosity demonstrates
 a breakdown of the SER below a crossover temperature of 700 K. These results  support the experimental evidence of a breakdown of SER in
 the similar compound GST
 inferred by Orava et al. from ultrafast DSC measurements \cite{orava}.
 This feature is the key to understand the origin of the high
 crystallization rate in phase change memories.
 In the set process of PCM, the amorphous phase is heated above T$_g$. Preliminary simulation results actually show that the glass transition  experiences
a hysteresis, the rapidly overheated amorphous
phase differing from the supercooled liquid in a narrow range of temperature above T$_g$. The overheated amorphous phase
displays a breakdown of SER as well,
although somehow less pronounced than in the supercooled liquid. This issue will be addressed in a future publication.
However, in the set process of PCM, the temperature can rise significantly above T$_g$ depending
on the details of the programming pulses.
 Indeed optimized set pulses with peak intensities equal to the reset pulse (leading to the melting of
the crystal) are  implemented in current
PCM devices \cite{borghi}. Under these conditions the crystallization occurs from the supercooled liquid phase.
  In spite
 of a large viscosity,  a high diffusivity is still possible in the supercooled liquid down to  temperatures very close to T$_g$
 because of the breakdown of SER. This allows for the coexistence  of a high diffusivity and a high driving force for crystallization that boost the crystallization speed at high supercooling.
 The crystallization of the supercooled liquid or of a highly mobile overheated amorphous phase might take place in a different
manner with respect to the crystallization of the amorphous phase at temperatures below T$_g$ which is of interest for data retention.
The conclusion we can draw is that the self-diffusion coefficient in the supercooled liquid regime close to T$_g$
can not be inferred from the viscosity and
the application of the SER.
Secondly, the measured Arrhenius behavior of the diffusivity and/or crystallization speed below T$_g$ can not be extrapolated above T$_g$ in the supercooled liquid. These results are of interest also for the refinement of the electrothermal modeling of PCM devices  \cite{redaelli}.
\vskip 1.0 truecm

\section{Acknowledgment}
 This work has been partially supported
 by Italian MURST through the program Prin08, by  the Cariplo Foundation through project MONADS
 and by Regione Lombardia and CILEA Consortium
through a LISA Initiative (Laboratory for Interdisciplinary Advanced
Simulation) 2011 grant [link: http://lisa.cilea.it ].
 We thankfully acknowledge the computational resources
 also by the ISCRA Initiative at Cineca (Casalecchio di Reno, Italy) and by CSCS (Manno, Switzerland).

\end{document}